\documentclass[12pt,preprint]{aastex}
\input epsf
\eqsecnum
\begin{document}
\title{A Redshift-Magnitude Relation for Non-Uniform Pressure Universes}

\author{Mariusz P. D\c{a}browski}

\affil{Institute of Physics, University of Szczecin, Wielkopolska 15, 70-451
Szczecin, Poland and Astronomy Centre, University of Sussex, Falmer, Brighton
BN1 9QH, U. K..}

\begin{abstract}
A redshift-magnitude relation for the two exact non-uniform pressure spherically
symmetric Stephani universes is presented. The Kristian-Sachs method expanding
the relativistic quantities in series is used, but only first order terms in
redshift $z$ are considered. The numerical results are given both for centrally
placed and non-centrally placed observers. In the former case the
redshift-magnitude relation does not depend on the direction in the sky and the
Friedman limit can be easily performed. It appears that the effect of spatial
dependence of pressure is similar to the effect of the deceleration parameter
in Friedman models. In the latter case the angular dependence of the relation
is important. This may serve as another possible explanation of the
noncompatibility of the theoretical curve of the redshift-magnitude relation
with observations for large redshift objects in the Friedman universe.
On the other hand comparing the magnitudes of equal redshifts objects in
different directions in the sky one can test the reliability of these models.
\end{abstract}


{\bf Journal Reference: The Astrophysical Journal {\bf 447}, 43 (1995).}

\section{INTRODUCTION}
\vspace{0.6cm}

Inhomogeneous models of the Universe have gradually become more popular among
cosmologists. Although we know quite a lot of different inhomogeneous solutions
(e.g. Kramer et al. 1980, Krasi\'nski 1993) the most famous and strongly investigated
models of this general class have been the spherically symmetric dust Tolman
universes (Tolman 1934, Bondi 1947, Bonnor 1974). Their properties have been
studied quite thoroughly by Hellaby and Lake (1984, 1985), Hellaby
(1987, 1988) and the observational relations for them were studied by Goicoechea and
Martin-Mirones (1986), Moffat and Tatarski (1992).

Recently, we have considered the global properties of the spherically symmetric
Ste\-pha\-ni universes (D\c{a}browski 1993). A couple of exact inhomogeneous
solutions have been found and the question arises how far from the real
Universe these solutions can be. The purpose of this paper is to give the
observational relations that could enable us to compare these solutions with
astronomical observations of galaxies and quasars.

Generally, in inhomogeneous models the density and (or) pressure depend on
spatial coordinates. The density in the Tolman models is non-uniform. By
this we mean that it depends on both the time and the radial coordinate. On the
other hand the pressure is uniform and it only depends on the time coordinate.
In the Stephani models it is quite different - the density is uniform and the
pressure is non-uniform. Although it seems to be easier to think about
non-uniform density in the Universe, the non-uniform pressure could still have
some motivation in inflationary cosmology, at least in the very early stages
of the Universe when the vacuum pressure plays an important role (Vilenkin
1985, Weinberg 1989, Linde 1994). The Stephani models have also been studied in
the context of thermodynamics by Sussman (1994), Quevedo and Sussman (1994a,b).

Some other possibilities of admitting inhomogeneities in the Universe have been
studied, among others, by Roeder (1975), Dyer (1979), Partovi and Mashhoon 
(1984). Some observational quantities for these models have also been found.

In this paper we shortly comment on the Stephani models in Section 2. In
Section 3 we present a formula for the redshift in Model I (MI) and Model II (MII)
which have been fully considered in the earlier paper (D\c{a}browski 1993).
In Section 4 we use the Kristian-Sachs method (Kristian and Sachs 1966) to
derive the redshift-magnitude
relations for MI and MII. In Section 5 we present numerical results for the
redshift-magnitude relation both for centrally and non-centrally placed
observers. In the Appendix A we write down the components of the null tangent
vector to zero geodesic equations and solve them in a couple of cases. In the
Appendix B we present a redshift-magnitude relation for a radial ray in the
general spherically symmetric Stephani universe.

\vspace{.6cm}
\section{The Models}
\vspace{.6cm}
The spherically symmetric Stephani metric is given by (Krasi\'{n}ski 1983,
D\c{a}browski 1993)
\begin{equation}
\label{STMET}
ds^2~=~-~D^2c^2dt^2~+~ \frac{R^2}{V^2} \left[dr^2~+~r^2 \left(d\theta^2~+~
\sin^2{\theta}d\varphi^2 \right) \right]~,
\end{equation}
where
\begin{eqnarray}
\label{VDK}
  V(t,r) & = & 1 + \frac{1}{4}kr^2~,\\
  D(t,r) & = & F \frac{R}{V} \left( \frac{V}{R} \right)^{\cdot}~,\\
    k(t) & = & \left( C^2(t)~-~\frac{1}{c^2F^2(t)} \right) ~R^2(t)~,
\end{eqnarray}
and $(\ldots)^{\cdot}~\equiv~\frac{\partial}{\partial t}$. In (2.1)-(2.4) we
have chosen the following units: the time t is taken in sMpc/km, the function
$R(t)$ (generalized scale factor)
is taken in megaparsecs or kilometers, the function $F(t)$ is in seconds, the
dimension of $C(t)$ is $km^{-1}$, $c = 3 \cdot 10^5$ km/s is the velocity of light
and all the other functions $r, \theta, \phi, D(t,r), V(t,r), k(t)$ are
dimensionless. The mass density and pressure are given by
\begin{eqnarray}
\label{ROPE}
\frac{8\pi G}{c^2} \rho(t) & = & 3C^2(t)~,\\
\frac{8\pi G}{c^4} p(t) & = & -~3C^2(t)~+~2C \dot{C} \frac{ \left( \frac{V}{R} \right)}
  { \left( \frac{V}{R} \right)^{\cdot}}~,
\end{eqnarray}
where $G$ is the gravitational constant. This means that the density is
uniform while the pressure is non-uniform
throughout the Stephani universe. The four-velocity of matter has only one
nonvanishing component
\begin{equation}
u_{t}~=~-~cD~,
\end{equation}
and the only nonvanishing component of the acceleration is
\begin{equation}
\dot{u}_{r}~=~c\frac{D_{,r}}{D}~,
\end{equation}
and $(\ldots)_{,r}~\equiv~\frac{\partial}{\partial r}$~.

In this paper we will be considering only two subcases of the model (2.1),
which possess the flat Friedman limit, so the comparison with the isotropic
data seems to be very easy for them. We will be called these subcases Model I
(MI) and Model II (MII), respectively. These models admit a Friedman-like time
coordinate
\begin{equation}
d\tau~=~-~\int F \frac{\dot{R}}{R} dt~,
\end{equation}
in which the expansion scalar is simply $\Theta~=~3H~=~3R_{,\tau}/R$.
For the Model I we have $C(t) = A R(t)$ with $A = $const. (D\c{a}browski 1993)
and
\begin{eqnarray}
\label{MI}
  k(\tau) & = & -~4\frac{a}{c^2}R(\tau)~,\\
  R(\tau) & = & a\tau^2~+~b\tau~+~d~,\\
  V(\tau,r) & = & 1~-~\frac{a}{c^2} \left( a\tau^2~+~b\tau~+~d \right) r^2~,\\
  \Delta & \equiv & 4ad~-~b^2~+~1~=~0~,
\end{eqnarray}
with a, b, d = const. and for the cosmic time $\tau$ taken in sMpc/km we have:
[a] = $km^2/(s^2Mpc)$, [b] = km/s and [c] = Mpc. For the Model II we have 
(Wesson and Ponce de Leon 1989, D\c{a}browski 1993)
\begin{eqnarray}
\label{MII}
  k(\tau) & = & -~\frac{\alpha \beta}{c^2} R(\tau)~,\\
  R(\tau) & = & \beta \tau^{\frac{2}{3}}~,\\
  V(\tau,r) & = & 1~-~\frac{1}{4c^2} \alpha \beta^2 \tau^{\frac{2}{3}} r^2~,
\end{eqnarray}
with $\alpha$, $\beta$ = const. with $[\alpha]$ = $(s/km)^{\frac{2}{3}}
Mpc^{-\frac{4}{3}}$  and $[\beta]$ = $(km/s)^{\frac{2}{3}} Mpc^{\frac{1}{3}}$.
Both models possess the Friedman limit; ($a \rightarrow 0$  for MI and
$\alpha \rightarrow 0$ for  MII). The common point
between MI and MII is that for them $ \left( \frac{k}{R} \right)_{,\tau} = 0$,
where $(\ldots)_{,\tau}~\equiv~ \frac{\partial}{\partial \tau}$
(cf. Appendix A) and $D~=~1/V$ (cf.(2.3)). The four-velocity and the
acceleration (2.7)-(2.8) read as
\begin{eqnarray}
  u_{\tau} & = & -~c\frac{1}{V}~,\\
  \dot{u}_{r} & = & -~c\frac{V_{,r}}{V}~.
\end{eqnarray}
The components of the vector tangent to zero geodesic are (see
Appendix A)
\begin{eqnarray}
\label{COM}
  k^{\tau} & = & \frac{V^{2}}{R}~,\\
  k^{r} & = & \pm \frac{V^{2}}{R^{2}} \sqrt{1~-~\frac{h^{2}}{r^{2}}}~,\\
  k^{\theta} & = & 0~,\\
  k^{\varphi} & = & h \frac{V^{2}}{R^{2}r^{2}}~,
\end{eqnarray}
where $h$ = const., and the plus sign in (2.20) applies to a ray moving away from
the centre, while the minus sign applies to a ray moving towards the centre.
The acceleration scalar for MI and MII respectively is
\begin{equation}
\dot{u}~\equiv \left( \dot{u}_{a} \dot{u}^{a} \right) ^{\frac{1}{2}}~=~
\frac{V_{,r}}{R}~=~\left\{ \begin{array}{l}
                             -~2\frac{a}{c^2}r ,\\
                             -~\frac{1}{2} \alpha \beta r ,
                           \end{array}
                   \right.\
\end{equation}
and it does not depend on the time coordinate at all. Also, it means that the
further away from the center $r=0$ is an observer, the larger acceleration he
subjects.
\vspace{.6cm}
\section{The Redshift}
\vspace{.6cm}
For any cosmological model the redshift is given by (Ehlers 1961, Ellis and
MacCallum 1970)
\begin{equation}
\label{RED}
1~+~z~=~\frac{ \left( u_{a}k^{a} \right)_{O}}{ \left( u_{a}k^{a} \right)_{G}}~,
\end{equation}
where index 'O' means that the quantities should be taken at the observer
position, while index 'G' means that the quantities should be taken at the
galaxy position. According to (3.1), (2.7) and (2.19)-(2.22) we have for MI
and MII respectively
\begin{equation}
1~+~z~= \left\{ \begin{array}{l}
                  \frac{ \left[ \frac{1~-~\frac{a}{c^2}(a\tau^2~+~b\tau~+~d)r^2}
                                     {a\tau^2~+~b\tau~+~d} \right]_{O}}
                       { \left[ \frac{1~-~\frac{a}{c^2}(a\tau^2~+~b\tau~+~d)r^2}
                                     {a\tau^2~+~b\tau~+~d} \right]_{G}}  ,\\

                  \frac{ \left[ \frac{1~-~\frac{1}{4}\alpha\beta^2\tau^\frac
                                     {2}{3}r^2}{\beta\tau^\frac{2}{3}}
                                     \right]_{O}}
                       { \left[ \frac{1~-~\frac{1}{4}\alpha\beta^2\tau^\frac
                                     {2}{3}r^2}{\beta\tau^\frac{2}{3}}
                                     \right]_{G}}  .
                \end{array}
        \right.\
\end{equation}
For corresponding flat Friedman limits
\begin{equation}
1~+~z~=~\left\{ \begin{array}{l}
                  \frac{(b\tau~+~d)_{G}}{(b\tau~+~d)_{O}}  ,\\
                  \frac{(\tau^\frac{2}{3})_{G}}{(\tau^\frac{2}{3})_{O}}  .
                \end{array}
        \right.\
\end{equation}
The first limit corresponds to the exotic equation of state model
$p~=~-~\frac{1}{3}\rho$ (Vilenkin 1985, D\c{a}browski and Stelmach 1989)
and the second corresponds to the dust model $p~=~0$.

From (3.2) one can easily notice that, for instance for the model MII, if the
observer is at the centre of
symmetry $r~=~0$ the redshift of a given galaxy lying away from the centre
is greater than in corresponding Friedman model (3.3). On the other hand, if
it is a galaxy at the centre $r~=~0$ and the observer is away from it, the
redshift of the galaxy measured by the observer is smaller than in the
corresponding Friedman model (3.3).

\vspace{.6cm}
\section{The Redshift-Magnitude Relation}
\vspace{.6cm}

The standard procedure in order to obtain the redshift-magnitude relation
in power series around the observer position and time is based on the
formalism given by Kristian and Sachs (1966) and Ellis and MacCallum 
(1970).
In this paper we assume the signature convention $(-,+,+,+)$
used by Ellis and MacCallum. The redshift-magnitude formula is given by
\begin{eqnarray}
\label{MBOL}
m_{bol} = M - 5\log_{10}{ \left( u_{a;b}K^{a}K^{b} \right)_{O}} + 5\log_{10}{cz}
\nonumber \\
+ \frac{5}{2} \left( \log_{10}{e} \right) \left\{ \left( 4 - \frac {u_{a;bc}K^{a}
K^{b}K^{c}}{ \left( u_{a;b}K^{a}K^{b} \right)^2} \right) z +
{\bf O} \left( z^2 \right) \right\}_{O}~,
\end{eqnarray}
where
\begin{eqnarray}
\label{OTH}
  u_{a;b} & = & \frac{1}{3} \Theta h_{ab} - \dot{u}_{a} u_{b}~,\\
  h_{ab} & \equiv & g_{ab} + u_{a}u_{b}~,\\
  K^{a} & \equiv & \frac{k^{a}}{u_{b}k^{b}}~,\\
  u_{a}u^{a} & = & - 1~.
\end{eqnarray}
Here $m_{bol}$ is the bolometric apparent magnitude, $M$ - absolute magnitude,
$\Theta$ - the expansion scalar, $\dot{u}_{a}$ - the acceleration vector,
$u_{a}$ - the four-velocity of matter, $k^{a}$ - null vector tangent to the
zero geodesic that connects galaxy and observer, $h_{ab}$ - the operator that
projects vectors onto spacelike hypersurfaces ($h_{ab}u^{b}=0$,
$K_{a}K^{b}=0$, $\dot{u}_{a}u^{a}=0$, $u_{a}K^{a}=1$). The projection of
$K^{a}$ onto the spatial hypersurfaces ortogonal to $u_{a}$ - a spatial unit
vector pointing in the observer direction of the source is
\begin{equation}
n^{a} = - u^{a} - K^{a} ,
\end{equation}
and
\begin{equation}
n^{a}n_{a} = 1 .
\end{equation}
From (4.2)-(4.4) we have
\begin{eqnarray}
\label{ABC}
u_{a;bc} = \frac{1}{3} \Theta_{,c} h_{ab} + \frac{1}{9} \Theta^2 \left( h_{ac}
u_{b} + h_{bc}u_{a} \right) - \dot{u}_{a;c} u_{b} \nonumber \\
- \frac{1}{3} \Theta \left( \dot{u}_{a}u_{b}u_{c} + u_{a}\dot{u}_{b}u_{c}
\right) - \frac{1}{3} \Theta \dot{u}_{a}h_{bc} + \dot{u}_{a} \dot{u}_{b} u_{c} ,
\end{eqnarray}
In order to calculate (4.1) we should use (4.3) and put it into (4.2) and (4.8)
i.e.
\begin{eqnarray}
\label{KK}
u_{a;b}K^{a}K^{b} & = & \frac{1}{3} \Theta - \dot{u}_{a}K^{a} ,\\
u_{a;bc}K^{a}K^{b}K^{c} & = & \frac{1}{3} \Theta_{,c} K^{c} + \frac{2}{9}
\Theta^2 - \Theta \dot{u}_{a} K^{a} - \dot{u}_{a;c}K^{a}K^{c} +
\dot{u}_{a} \dot{u}_{b} u_{c}  ,
\end{eqnarray}
what according to (2.7) and (2.8) gives
\begin{eqnarray}
\label{KRKT}
u_{a;b}K^{a}K^{b} & = & \frac{1}{3} \Theta + c\frac{D_{,r}}{D^2}
\frac{k^{r}}{k^{t}} ,\\
u_{a;bc}K^{a}K^{b}K^{c} & = & - \frac{1}{3} \frac{\dot{\Theta}}{D} +
\frac{2}{9} \Theta^2 + \frac{c}{D^2} \frac{k^{r}}{k^{t}} \times \nonumber \\
&   & \left\{4\Theta D_{,r}
- \left( \frac{D_{,r}}{D} \right)_{,t}- c\left[ \left(
\frac{D_{,r}}{D} \right)_{,r} + \frac{V_{,r}}{V} \frac{D_{,r}}{D} -
\left( \frac{D_{,r}}{D} \right)^2 \right]\frac{k^{r}}{k^{t}} \right\} .
\end{eqnarray}
For models MI and MII, according to (2.17)-(2.22) we have
\begin{eqnarray}
\label{M1M2}
u_{a;b}K^{a}K^{b} & = & \frac{R_{,\tau}}{R} \mp c\frac{V_{,r}}{R}
\sqrt{1 - \frac{h^2}{r^2}} ,\\
u_{a;bc}K^{a}K^{b}K^{c} & = & 2 \left( \frac{R_{,\tau}}{R}
\right)^2 - \left( \frac{R_{,\tau}}{R} \right)_{,\tau}V \pm
c \frac{V^2}{R} \left[ \left( \frac{V_{,r}}{V} \right)_{,\tau}
- 12 \frac{R_{,\tau}}{R} \frac{V_{,r}}{V^2}  \right] \nonumber \\
&   & \times  \sqrt{1 - \frac{h^2}{r^2}}
+ c^2 \left[ 2 \left( \frac{V_{,r}}{V} \right)^2 + \left( \frac{V_{,r}}{V}
\right)_{,r} \right] \frac{V^2}{R^2} \left( 1 - \frac{h^2}{r^2} \right) .
\end{eqnarray}
Finally,
\begin{eqnarray}
u_{a;b}K^{a}K^{b} & = & \frac{R_{,\tau}}{R} \mp  2\frac{a}{c}r \sqrt{1 - \frac{h^2}
{r^2}}  ,\\
u_{a;bc}K^{a}K^{b}K^{c} & = & 2 \left( \frac{R_{,\tau}}{R} \right)^2
- \left( \frac{R_{,\tau}}{R} \right)_{,\tau} \left( 1 - \frac{a}{c^2}Rr^2
\right) \nonumber \\
&   & \mp  22\frac{a}{c}r \frac{R_{,\tau}}{R} \sqrt{1 - \frac{h^2}{r^2}}
- \frac{2a}{R} \left( 1 - 3\frac{a}{c^2}Rr^2 \right) \left( 1 - \frac{h^2}{r^2} \right) ,
\end{eqnarray}
for MI, and
\begin{eqnarray}
u_{a;b}K^{a}K^{b} & = & \frac{R_{,\tau}}{R} \mp \frac{1}{2} c \alpha \beta r
\sqrt{1 - \frac{h^2}{r^2}}  ,\\
u_{a;bc}K^{a}K^{b}K^{c} & = & 2 \left( \frac{R_{,\tau}}{R} \right)^2
- \left( \frac{R_{,\tau}}{R} \right)_{,\tau} \left( 1 - \frac{1}{4}\alpha\beta
Rr^2 \right) \mp \nonumber \\ &   &
c \frac{11}{2} \alpha\beta r \frac{R_{,\tau}}{R} \sqrt{1 - \frac{h^2}{r^2}}
- \frac{c^2}{2} \frac{\alpha\beta}{R}\left( 1 - \frac{3}{4} \alpha
\beta Rr^2 \right) \left( 1 - \frac{h^2}{r^2} \right),
\end{eqnarray}
for MII.

The formulas (4.15)-(4.18) taken at the observer position (index "O") have to be
inserted into the magnitude-redshift relation (4.1) to compare MI and MII with
current observational data. However, from the point of view of observations the
more useful form of relation (4.1) can be obtained by applying a spatial unit
vector $n^{a}$ given by (4.6)-(4.7). Its radial component in the natural
orthonormal basis associated with the metric (2.1) at the observer
position is defined as
\begin{equation}
\left( n^{r} \right)_{O} = \cos{\phi} = \pm \left( \sqrt{1 -
\frac{h^2}{r^2}} \right)_{O} ,
\end{equation}
where $\phi$ is the angle between the direction of observation and the direction
defined by the observer and the centre. One can easily notice that if $0 < \phi <
\frac{\pi}{2}$ and $\frac{3\pi}{2} < \phi < 2\pi$, then $n^{r} > 0$ and the ray reaching the observer emitted by a
galaxy goes away from the centre (which corresponds to the plus sign in (2.20)),
and if $\frac{\pi}{2} < \phi < \frac{3\pi}{2}$ , $n^{r} < 0$ and the ray goes towards the
centre (minus sign in (2.20)).

In terms of the spatial vector $n^{a}$ the equations (4.9)-(4.10) are
\begin{eqnarray}
u_{a;b}K^{a}K^{b} & = & \frac{1}{3} \Theta - \dot{u}_{r}n^{r} ,\\
u_{a;bc}K^{a}K^{b}K^{c} & = & - \frac{1}{3} \frac{\dot{\Theta}}{D}
+ \frac{2}{9} \Theta^2 + \left( 4\Theta \dot{u}_{r} -
\frac{\dot{u}_{r,t}}{D} \right) n^{r} \nonumber \\
            &   & + \left[ \left( \dot{u}_{r}
\right)^2 - \dot{u}_{r,r} - \frac{V_{,r}}{V} \dot{u}_{r} \right] \left( n^{r}
\right)^2 ,
\end{eqnarray}
and, in turn, for models MI and MII $D = 1/V$, $\dot{u}_{r} = - V_{,r}/V$ , so
\begin{eqnarray}
u_{a;b}K^{a}K^{b} & = & H - c \frac{V_{,r}}{R} \cos{\phi}  ,\\
u_{a;bc}K^{a}K^{b}K^{c} & = & - \dot{H} V + 2H^2 + c \left[ V \left( \frac{V_{,r}}
{V} \right)_{,\tau} - 12H \frac{V_{,r}}{V} \right] \frac{V}{R} \cos{\phi}
\nonumber \\     &   &
+ c^2 \left[ 2 \left( \frac{V_{,r}}{V} \right)^2 + \left( \frac{V_{,r}}{V} \right)
_{,r} \right] \frac{V^2}{R^2} \cos^2{\phi}  ,
\end{eqnarray}
where all quantities should be taken at the observer position (index "O" - see
(4.1)). For comparison, different values of the angle $\phi$ in (4.12)-(4.13)
correspond to different values of the parameter h in (4.13)-(4.14).
Finally,
\begin{eqnarray}
u_{a;b}K^{a}K^{b} & = & \frac{R_{,\tau}}{R} + 2\frac{a}{c}r \cos{\phi}    ,\\
u_{a;bc}K^{a}K^{b}K^{c} & = & - \left( \frac{R_{,\tau}}{R} \right)_{,\tau}
\left( 1 - \frac{a}{c^2}Rr^2 \right) + 2 \left( \frac{R_{,\tau}}{R} \right)^2 \nonumber \\
            &   & + 22\frac{a}{c}r \frac{R_{,\tau}}{R} \cos{\phi}
- 2ar \frac{1 - 3\frac{a}{c^2}Rr^2}{R} \cos^2{\phi}  ,
\end{eqnarray}
for MI, and
\begin{eqnarray}
u_{a;b}K^{a}K^{b} & = & \frac{R_{,\tau}}{R} + \frac{c}{2}\alpha\beta r
\cos{\phi}  ,\\
u_{a;bc}K^{a}K^{b}K^{c} & = & - \left( \frac{R_{,\tau}}{R} \right)_{,\tau}
\left( 1 - \frac{1}{4}\alpha \beta Rr^2 \right) + 2 \left( \frac{R_{,\tau}}{R} \right)^2
+ \nonumber \\ &   &
\frac{11}{2} c \alpha\beta r \frac{R_{,\tau}}{R}\cos{\phi} -
\frac{1}{2} c^2 \alpha\beta  \frac{1 - \frac{3}{4}\alpha\beta Rr^2}{R}
\cos^2{\phi} ,
\end{eqnarray}
for MII.

\vspace{.6cm}
\section{Numerical Results}
\vspace{.6cm}
In this section we plot the redshift-magnitude relation (4.1) for models MI and
MII with some specific values of their parameters chosen. One has to remember
that in this paper we neglect all the non-linear terms in z (cf. (4.1)).
We consider two basic cases:
\vspace{.5cm}
\begin{center}
{\em a) centrally placed observers}
\end{center}
\vspace{.5cm}
For centrally placed observers the radial coordinate $r_{0} = 0$ and $h = 0$ in
(2.19)-(2.22). In such a case it is much more convenient to use the relations
(4.15)-(4.18) instead of (4.24)-(4.27) since we can find easily the
correspondance with FRW universes by assuming Stephani parameters approaching
zero.

For the model MI from (4.15) and (4.16) we have
\begin{eqnarray}
u_{a;b}K^{a}K^{b} \mid_{0} & = & \frac{R_{,\tau}}{R} \mid_{0}  ,\\
u_{a;bc}K^{a}K^{b}K^{c} \mid_{0} & = & 2 \left( \frac{R_{,\tau}}{R} \right)^2
\mid_{0} - \left( \frac{R_{,\tau}}{R} \right)_{,\tau} \mid_{0}
- \frac{2a}{R} \mid_{0}   ,
\end{eqnarray}
and from (4.1)
\begin{eqnarray}
\label{MBOL1}
m_{bol} = M - 5\log_{10}{ \bar{H}_{0}} + 5\log_{10}{cz}
+ 1.086 \left( 1 - 2\bar{q_{0}} \right) z   ,
\end{eqnarray}
where $\bar{H_{0}}$ and $\bar{q_{0}}$ are of the same form as in the Friedman
universe i.e.
\begin{eqnarray}
\bar{H_{0}} & \equiv & \frac{R_{,\tau}}{R} \mid_{0}   ,\\
\bar{q_{0}} & \equiv & - \frac{R R_{,\tau \tau}}{R^2_{,\tau}} \mid_{0} =
- \frac{2aR}{R^2_{,\tau}} \mid_{0}   .
\end{eqnarray}
Finally, for units taken in megaparsecs we have
\begin{eqnarray}
\label{MB1}
m_{bol} = M + 25 + 5\log_{10}{ \left[cz \left(\frac{a\tau_{0}^2 + b\tau_{0}
+ d}{2a\tau_{0} + b}\right)\right]} +
1.086 \left[ 1 + 2a \frac{\left( a\tau_{0}^2 + b\tau_{0} + d
\right)}{\left(2a\tau + b \right)^2} \right] z   .
\end{eqnarray}
In the Friedman limit $a \rightarrow 0$ we have
\begin{eqnarray}
\label{MBF}
m_{bol} = M + 25 + 5\log_{10} {\left( \tau + \frac{d}{b} \right)_{0}}
+ 5\log_{10}{(cz)} + 1.086 z    ,
\end{eqnarray}
so $\bar{q}_{0} = 0$ in this case, which, provided $d = 0$, is in agreement with
Eq.(41) of the paper by D\c{a}browski and Stelmach (1989).

\begin{figure}[t]
\centering
\leavevmode\epsfysize=14cm \epsfbox{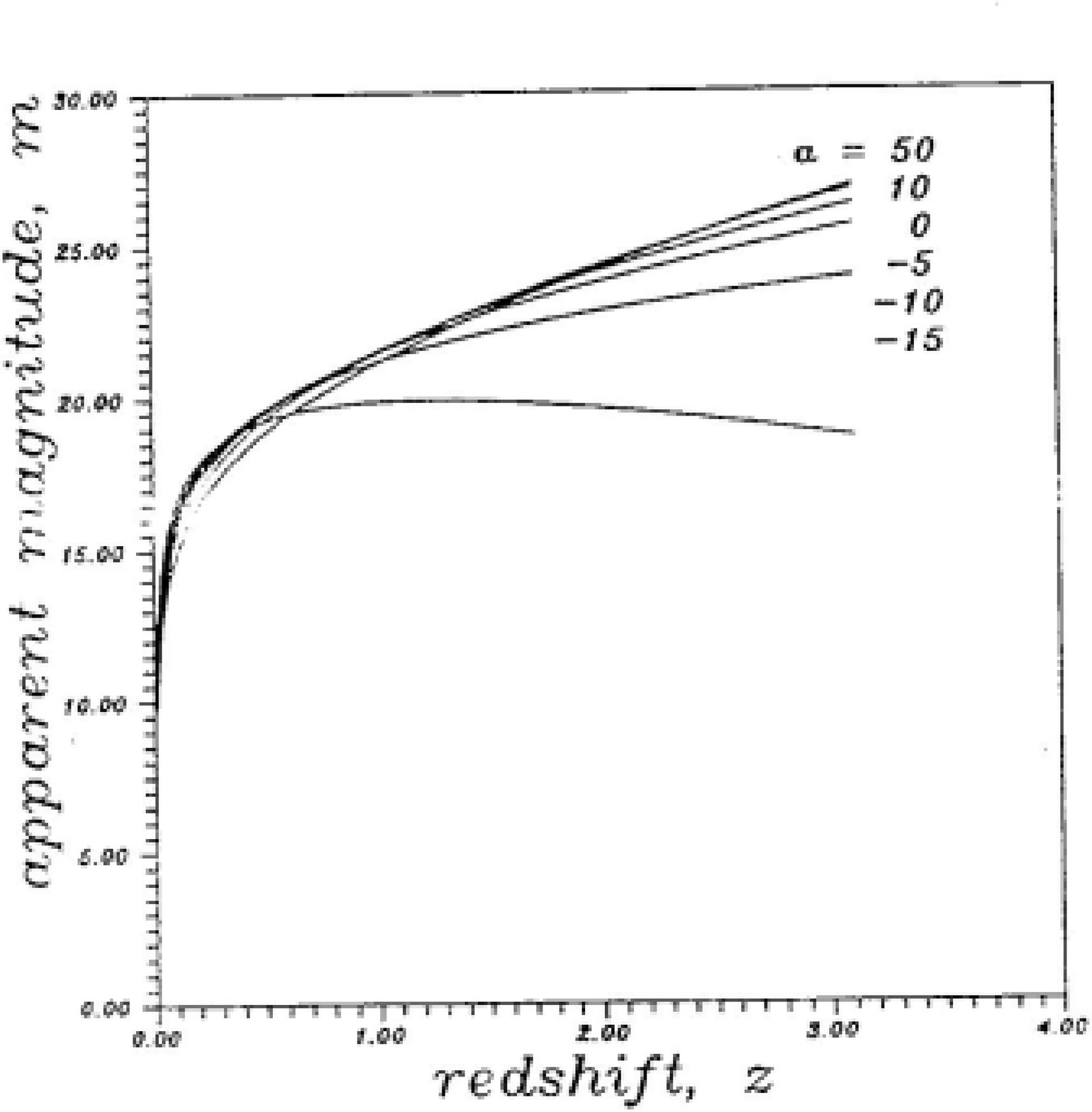}\\
\caption[]
{A plot of the redshift-magnitude relation for the model MI given by the
formulas (5.6)-(5.7). We have chosen a = -15, -10, -5, 0, 10, 50 $km^2/s^2Mpc$,
$b = 1 km/s$, d = 0 with $\tau_{0} = 0.02$ $sMpc/km$ and $M = - 23.5$. The
effect of non-uniform pressure is similar to the effect of spatial curvature
(expressed in terms of the deceleration parameter $q_{0}$) in FRW models.}
\label{fig1}
\end{figure}


The redshift-magnitude relations for the cases (5.6)-(5.7) are plotted in
\ref{fig1}. We have chosen the age of the universe $\tau_{0} = 0.02$ [sMpc/km], $M =
-23.5$ and the constants $a, b, d$ are taken in the following units: [a] =
$km^2/(s^2Mpc)$, [b] = km/s and [d] = Mpc. From Fig.1 one can
conclude that the effect of acceleration (2.23) is similar to the effect of
curvature (expressed in terms of deceleration parameter $q_{0}$) in the
Friedman models. Although we plotted the relation for arbitrary $z$ we have to
remember that we dropped the terms ${\bf O}\left( z^2 \right)$, so the results
can be slightly different for large redshifts.

For the model MII from (4.17) and (4.18) we have
\begin{eqnarray}
u_{a;b}K^{a}K^{b} \mid_{0} & = & \frac{R_{,\tau}}{R} \mid_{0}  ,\\
u_{a;bc}K^{a}K^{b}K^{c} \mid_{0} & = & 2 \left( \frac{R_{,\tau}}{R} \right)^2
\mid_{0} - \left( \frac{R_{,\tau}}{R} \right)_{,\tau} \mid_{0}
- \frac{1}{2} c^2 \frac{\alpha \beta}{R} \mid_{0}   ,
\end{eqnarray}
and from (4.1)
\begin{eqnarray}
\label{MBOL2}
m_{bol} = M - 5\log_{10} {\tilde{H}_{0}} + 5\log_{10}{cz}
+ 1.086 \left( 1 - \tilde{q_{0}} + \frac{9}{8} c^2 \alpha \tau_{0}^{\frac{4}{3}}
\right) z    ,
\end{eqnarray}
where
\begin{eqnarray}
\tilde{H_{0}} & \equiv & \frac{R_{,\tau}}{R} \mid_{0} \hspace{1.5mm}=
\frac{2}{3\tau_{0}}  ,\\
\tilde{q_{0}} & \equiv & - \frac{R R_{,\tau \tau}}{R^2_{,\tau}} \mid_{0}
\hspace{1.5mm}= \frac{1}{2}   .
\end{eqnarray}
are the Friedman values of the Hubble constant and the
deceleration parameter. In the Friedman limit $\alpha \rightarrow 0$ we
have
\begin{eqnarray}
\label{MBOLFr}
m_{bol} = M + 25 - 5\log_{10}{ \left( \frac{2}{3\tau} \right)_{0} }
+ 5\log_{10}{(cz)} + 0.543 z    ,
\end{eqnarray}
which corresponds to Eq.(41) of D\c{a}browski and Stelmach (1989)
for the flat model $\tilde{q_{0}} = \frac{1}{2}$.

In the model MII we have only one free parameter $\alpha$ which describes
inhomogeneity of the model. It can both be negative and positive and its effect
on the redshift-magnitude relation is similar to the effect of curvature in FRW
case (i.e. the values of $q_{0}$). In Fig.2 we plot suitable relations for
different $\alpha$ given in units $(km/sMpc)^{-\frac{4}{3}}$ , $M = - 23.5$ and
$\tau_{0}^{-1} = 75$ km/(sMpc). The main difference between the models MI and MII
is that in MII the curves for different $\alpha$ become separated for redshifts
approximately larger than 0.3 while in MI they differ almost from the
beginning. It is the result of the fact that in the latter model the constants
a, b, d effects the generalized Hubble constant (5.4) more strongly.

\begin{figure}[t]
\centering
\leavevmode\epsfysize=14cm \epsfbox{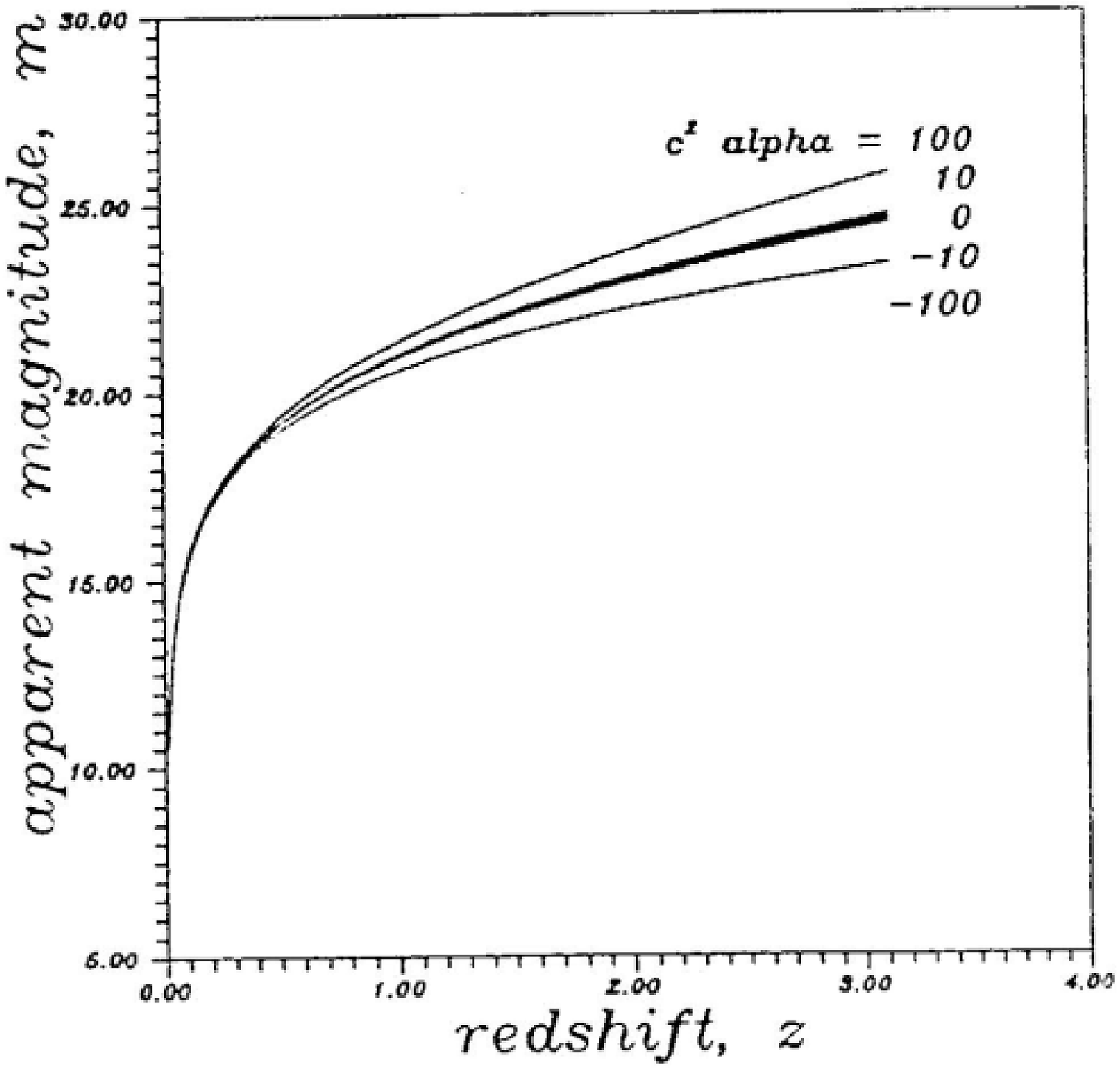}\\
\caption[]
{A plot of the redshift-magnitude relation for the model MII given by the
relation (5.10). Here $\alpha c^2 = 0, \pm 10, \pm 100 (km/sMpc)^{-\frac{4}{3}}$,
$\tau_{0}^{-1} = 75 km/sMpc$ and $M = - 23.5$. The effect of non-uniform
pressure is similar to the effect of spatial curvature (expressed in terms of
the deceleration parameter $q_{0}$) in FRW models.}
\label{fig2}
\end{figure}

\vspace{.5cm}
\begin{center}
{\em b) non-centrally placed observers}
\end{center}
\vspace{.5cm}

For non-centrally placed observers $r_{0} \neq 0$ and $h \neq 0$ in
(2.19)-(2.22) and the redshift-magnitude relation depends on the direction of a
galaxy in the observer sky. According to (4.26)-(4.27) the relation (4.1) for
MII reads as
\begin{eqnarray}
m_{bol} = M + 25 - 5\log_{10}{\left[ \left(\frac{R_{,\tau}}{R}\right)_{0} +
\frac{1}{2} c \frac{\alpha\beta}{c} r_{0} \cos{\phi} \right]} +
5\log_{10}{(cz)} + 4.344 z + 1.086 z \times \nonumber \\
\left[ \frac{ - 2 \left(\frac{R_{,\tau}}{R}\right)^2 +
\left(\frac{R_{,\tau}}{R} \right)_{,\tau}
\left(1 - \frac{\alpha\beta}{4} Rr_{0}^2 \right)
- \frac{11}{2} c \alpha\beta
\frac{R_{,\tau}}{R} r_{0} \cos{\phi} + c^2 \frac{\alpha\beta}{2R} \left(
1 - \frac{3}{4} \alpha\beta R r_{0}^2 \right) \cos^2{\phi}}
{\left( \frac{R_{,\tau}}{R} + \frac{1}{2} c \alpha\beta r_{0} \cos{\phi}
\right)^2} \right] \nonumber
\end{eqnarray}
or with $R(\tau)$ given explicitly by (2.15)
\begin{eqnarray}
m_{bol} = M + 25 + 5\log_{10}{\left[ \frac{cz}{\frac{2}{3} \frac{1}{\tau_{0}}
+ \frac{1}{2} c \alpha \beta r_{0} \cos{\phi}} \right]} + 1.086 z \times \nonumber \\
\left[ \frac{ \frac{2}{9} \frac{1}{\tau_{0}^2} \left( 1 + \frac{3}{4}
\alpha \beta^2 \tau_{0}^{\frac{2}{3}} r_{0}^2 \right) -
c \alpha \beta \frac{r_{0}}{\tau_{0}} \cos{\phi} +
\frac{1}{2} c^2 \alpha \tau_{0}^{-\frac{2}{3}} \left(
1 - \frac{5}{4} \alpha\beta^2 \tau_{0}^{\frac{2}{3}} r_{0}^2
\right) \cos^2{\phi}}
{\left( \frac{2}{3} \frac{1}{\tau_{0}} + \frac{1}{2} c \alpha\beta
r_{0} \cos{\phi} \right)^2} \right]    .
\end{eqnarray}

In Figures 3-5 we plot the dependence of the redshift-magnitude relation (5.14)
on the direction of the source in the sky and the distance from the center of
symmetry $r_{0}$. We fix the redshift of the source to be z = 0.1, 0.5 and 1.0
correspondingly and the other parameters are the following: $\alpha c^2 = 10
(km/sMpc)^{-\frac{4}{3}}, \beta = 1.1 \cdot 10^5 (km/s)^{\frac{2}{3}}
Mpc^{\frac{1}{3}}, \tau_{0}^{-1} = 75  km/(sMpc), -1 < \cos{\phi} < 1$. One can
easily notice that unlikely to the case of the centrally placed observer the
constant $\beta$ effects the redshift-magnitude relation as well. In fact, it is
the constant which appears in the flat dust-filled Friedman limit of the
Stephani Universe $\alpha \rightarrow 0$ (cf.(2.15)) and it has the same
dimension. In the Friedman model its value is $\beta = 1.1 \cdot 10^5
(km/s)^{\frac{2}{3}} Mpc^{\frac{1}{3}}$ for $\tau_{0}^{-1} = 75 km/(sMpc)$ i.e. $H_{0} = 50
km/(sMpc)$. From the Figures 3-5 we can see that for strongly non-centrally placed
observers the redshift-magnitude relation (for fixed $r_{0}$) becomes more
and more asymmetric. The smallest apparent magnitude is for the galaxies for
which the angle $\phi = \pi$ ($\cos{\phi} = - 1$) and they are just behind the
centre of symmetry with respect to the observer. The largest apparent magnitude
is achieved for $\phi = 0$ ($\cos{\phi} = 1$) and the galaxies are in front of
the centre of symmetry with respect to the observer. For the sake of comparison
(although we cannot take the limit $\alpha \rightarrow 0$ without taking
$r_{0} = 0$) we also draw the Friedman values of the apparent magnitude which
is not dependent on the angle $\phi$.

For the non-centrally placed observer one can think about a modification of the
centrally placed picture given in Fig. 2 in such a way that for each value of
the redshift $z$ we draw an "error" bar which range is given by the appropriate
values of $m(z)$ given by the non-centrally placed picture of figures
similar to Figures 3-5. This suggests one of the ways to explain the well-known
noncompatibility of the theoretical curve with the observational data for
large redshift galaxies in the Friedman universe. However, we emphasize that we
do not take any evolutionary effects into account here.

\begin{figure}[t]
\centering
\leavevmode\epsfysize=14cm \epsfbox{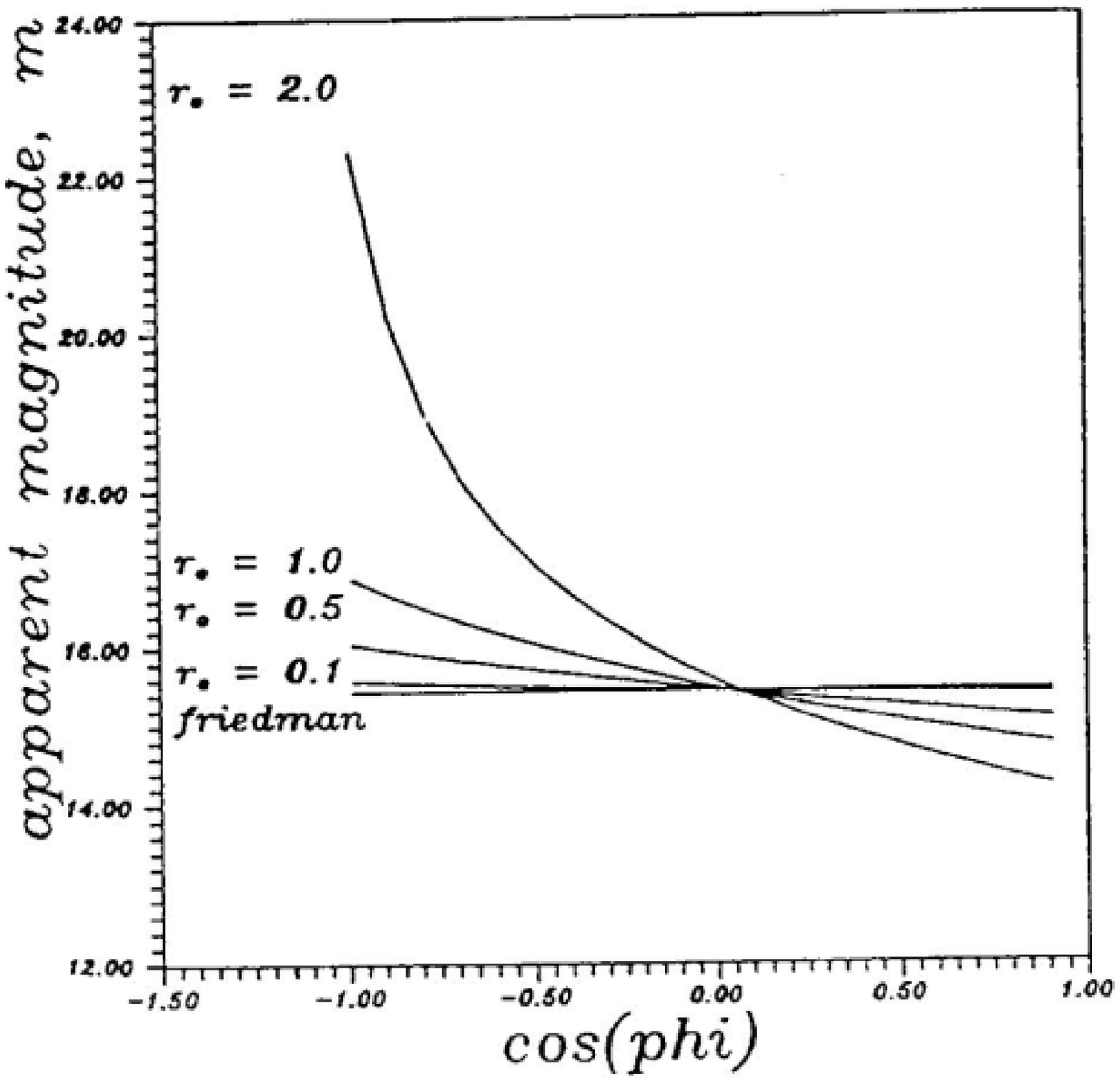}\\
\caption[]
{A plot of the dependence of the apparent magnitude on the direction in
the sky for the model MII according to the formula (5.14). We fix the redshift
of a galaxy to be
$z = 0.1$ and $\alpha c^2 = 100 (km/sMpc)^{-\frac{4}{3}}$, $\beta = 1.1 \cdot 10^5
(km/s)^{\frac{2}{3}} Mpc^{\frac{1}{3}}$, $\tau_{0}^{-1} = 75 km/(sMpc)$,
$-1 < \cos{\phi} < 1$
and $r_{0} = 0.1, 0.5, 1.0, 2.0$. If $\cos{\phi} = -1$ galaxies are just behind the
centre of symmetry with respect to the observer and the apparent magnitude is
small. If $\cos{\phi} = 1$ galaxies are in front of the centre of
symmetry and the apparent magnitude is large. The symmetric Friedman
value is given as well.}
\label{fig3}
\end{figure}

\begin{figure}[t]
\centering
\leavevmode\epsfysize=14cm \epsfbox{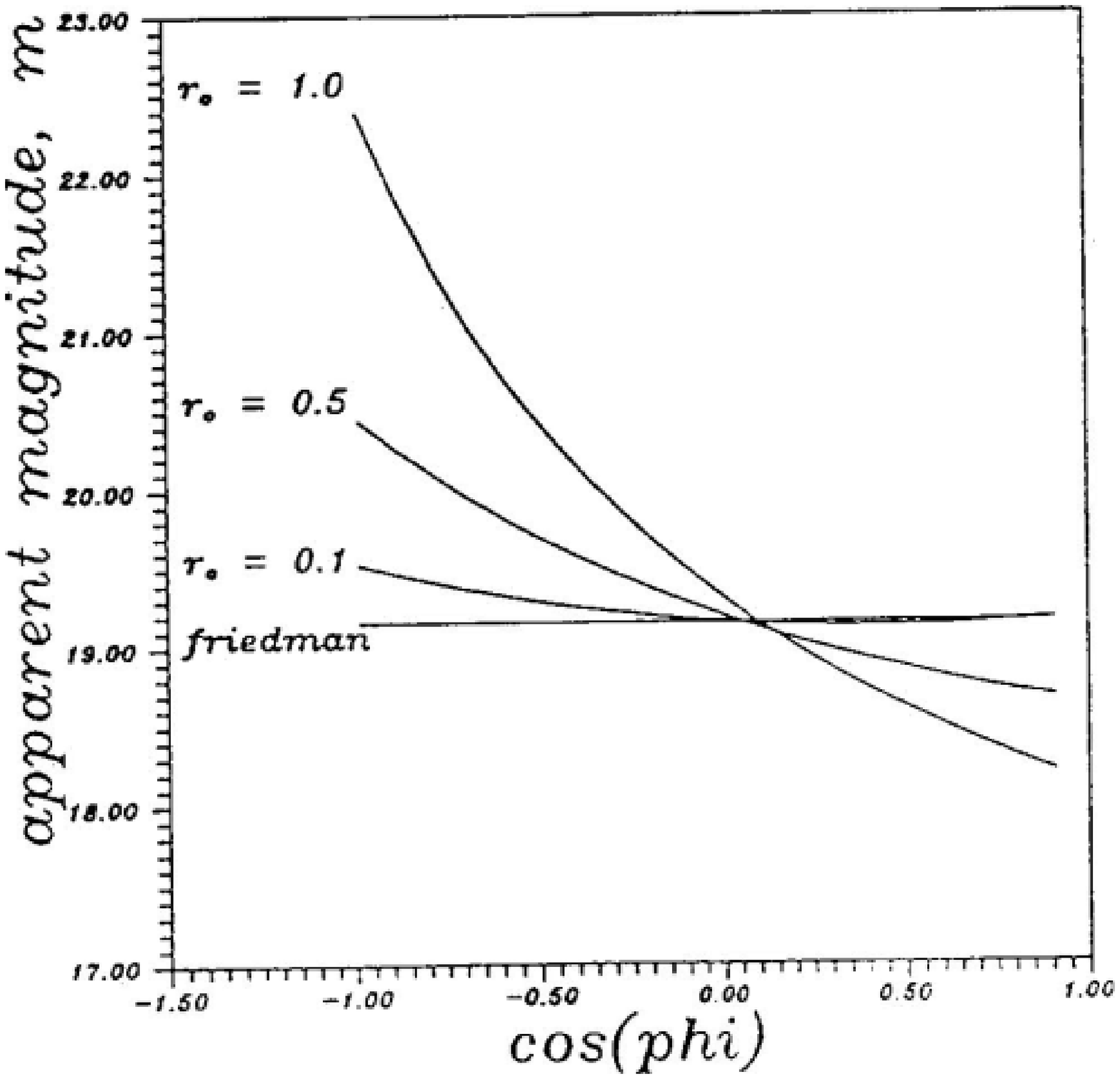}\\
\caption[]
{A plot of the dependence of the apparent magnitude on the direction in
the sky for the model MII according to the formula (5.14). We fix the redshift of a galaxy to be
$z = 0.5$ and $\alpha c^2 = 100 (km/sMpc)^{-\frac{4}{3}}, \beta = 1.1 \cdot 10^5
(km/s)^{\frac{2}{3}} Mpc^{\frac{1}{3}}, \tau_{0}^{-1} = 75 km/(sMpc),
-1 < \cos{\phi} < 1$ and $r_{0} = 0.1, 0.5, 1.0$.}
\label{fig4}
\end{figure}

\begin{figure}[t]
\centering
\leavevmode\epsfysize=14cm \epsfbox{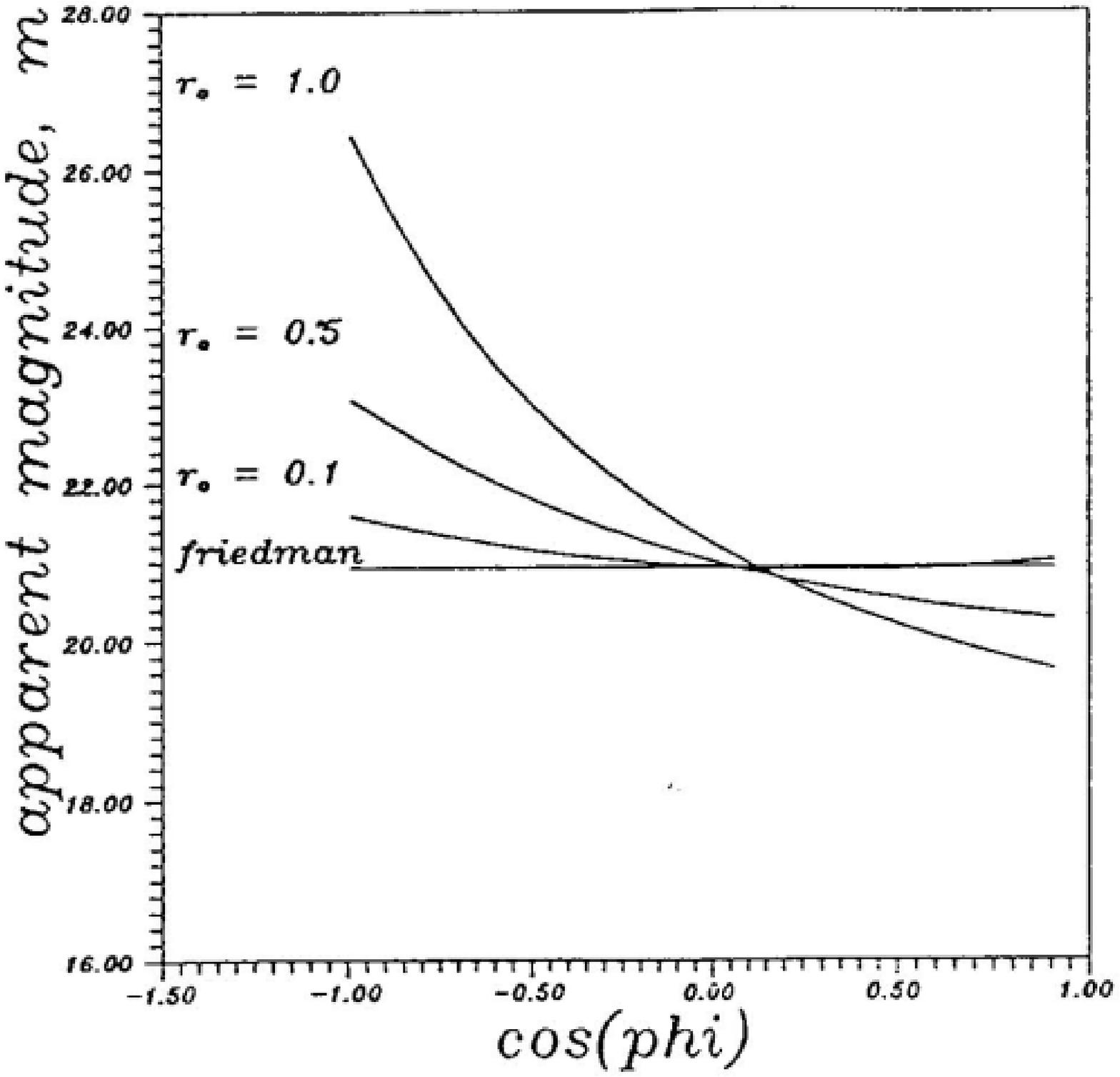}\\
\caption[]
{A plot of the dependence of the apparent magnitude on the direction in
the sky for the model MII according to the formula (5.14). We fix the redshift of a galaxy to be
$z = 1.0$ and $\alpha c^2 = 100 (km/sMpc)^{-\frac{4}{3}}, \beta = 1.1 \cdot 10^5
(km/s)^{\frac{2}{3}} Mpc^{\frac{1}{3}}, \tau_{0}^{-1} = 75 km/(sMpc),
-1 < \cos{\phi} < 1$ and $r_{0} = 0.1, 0.5, 1.0$.}
\label{fig5}
\end{figure}

\vspace{.6cm}
\section{Discussion}
\vspace{.6cm}

The comparison of the models MI and MII with astronomical data requires a
couple of conditions which have to be satisfied. For small redshift objects
there should not be any problem because our models deviate in a very clear
way from the flat Friedman universes but some difficulty might be related
to a position of the centre of symmetry from the observer (cf. Goicoechea and
Martin-Mirones 1987). Also, for some choices of the Stephani parameters
the singularities of pressure may appear (cf. D\c{a}browski 1993).
For large redshift galaxies and quasars the problem appears since the
Kristian-Sachs method of Section 4 generally valids if suitable series for
observational quantities are convergent. It seems to happen in the cases
considered in this paper, but in general it might not be so. Of course we have
still some freedom of a choice of functions $R(\tau)$ and $V(\tau,r)$
(cf. (2.2)-(2.4)) in order to make the series convergent.
On the other hand for large redshift objects the evolutionary effects ought to
be taken into account.

Of course the full information about the series can be obtained by
calculating the second and higher order corrections to m(z) in (4.1), but the
result contains the second and higher covariant derivatives of (4.2) (i.e.
$u_{a;bcd}$ - see Ellis and MacCallum (1970)) and the final formula is much more
complicated than the formulas (4.24)-(4.27). We decided to skip this calculations
in this paper and present them, if necessary, later.

The redshift formula (3.2) is always fulfilled, but with half of the quantities
taken at a galaxy position, which means that it is practically useless for
observational verification.

\appendix
\vspace{.6cm}
\section{Tangent vector to zero geodesic for Stephani models}
\vspace{.6cm}
If $k^{a} = \frac{dx^{a}}{ds}$, where $x^{a} = (t,r,\theta,\varphi)$ and s -
a parameter, is the null tangent vector to zero geodesic connecting observer
(index "O") and galaxy (index "G"), then the geometric optics equations for
the unknown components $k^{t}$ and $k^{r}$ in the spherically symmetric
Stephani universe (2.1) are
\begin{eqnarray}
F^2 \left[ \left( \frac{V}{R} \right)^{\cdot} \right]^2 \left( k^{t} \right)^2
= \left( k^{r} \right)^2 + h^2 \frac{V^4}{R^4r^2}  ,\\
\dot{k}^{t} k^{t} + k^{t}_{,r} k^{r} + 2 \left( \frac{\dot{R}}{R} -
\frac{\dot{V}}{V} \right) \left( k^{t} \right)^2 + \left[ \frac{\dot{F}}{F} +
\frac{ \left( \frac{V}{R} \right)^{\cdot\cdot}}{ \left( \frac{V}{R}
\right)^{\cdot}} \right] \left( k^{t} \right)^2 \nonumber \\
- 2 \left[ \frac{V_{,r}}{V} - \frac{ \left( \frac{V}{R} \right)^{\cdot}_{,r}}
{ \left( \frac{V}{R} \right)^{\cdot}} \right] k^{t}k^{r} = 0  ,\\
\dot{k}^{r} k^{t} + k^{r}_{,r} k^{r} + 2 \left( \frac{\dot{R}}{R} -
\frac{\dot{V}}{V} \right) k^{t}k^{r} - 2 \left[ \frac{V_{,r}}{V} - \frac{
\left( \frac{V}{R} \right)^{\cdot}_{,r}}{ \left( \frac{V}{R} \right)^{\cdot}}
\right] \left( k^{r} \right)^2 \nonumber \\
- h^2 \frac{V^4}{R^4r^2} \left( \frac{1}{r} -
\frac{V_{,r}}{V} \right) = 0  ,
\end{eqnarray}
while $k^{\theta} = 0$ and
\begin{equation}
k^{\varphi} = h \frac{V^2}{R^2r^2}  ,
\end{equation}
The easiest solutions of (A.1)-(A.4) are (D\c{a}browski 1993): \\
a) if $\left( \frac{V}{R} \right)^{\cdot\cdot} = 0$ i.e. $k(t) = \left( ct + d
\right) R(t)$ and $R(t) = \left( at + b \right)^{-1}$ ,$a, b, c, d = const.$,
then
\begin{eqnarray}
k^{t} & = & \frac{V^2}{FR^2} \frac{1}{ \left[ \left( \frac{V}{R}
\right)^{\cdot} \right]^2}  ,\\
k^{r} & = & \pm \frac{V^2}{R^2} \sqrt{ \frac{1}{ \left[ \left( \frac{V}{R}
\right)^{\cdot} \right]^2} - \frac{h^2}{r^2}}  .
\end{eqnarray}
b) if $\left( \frac{V}{R} \right)^{\cdot}_{,r} = 0$ i.e. $\left( \frac{k}{R}
\right)^{\cdot} = 0$, then
\begin{eqnarray}
k^{t} & = & \frac{V^2}{FR^2} \frac{1}{ \left( \frac{V}{R} \right)
^{\cdot}}  ,\\
k^{r} & = & \pm \frac{V^2}{R^2} \sqrt{ 1 - \frac{h^2}{r^2}}  .
\end{eqnarray}
If we use the Friedman-like time coordinate (2.9) we have to change the
derivative with respect to t into the derivative with respect to $\tau$ in
(A.1)-(A.8).
\vspace{.6cm}
\section{A redshift-magnitude relation for a radial ray}
\vspace{.6cm}
From (A.1) one can easily conclude that for moving radially towards or away
from the centre ray $\left( h = 0 \right)$ the ratio of components $k^{t}$ and
$k^{r}$ of the tangent vector is given by
\begin{equation}
\frac{k^{r}}{k^{t}} = \pm  D \frac{V}{R} = \pm \left( \frac{V}{R}
 \right)^{\cdot} ,
\end{equation}
and $k^{ \theta} = k^{ \varphi} = 0$.
This might be useful for calculating the redshift-magnitude relation.
From (4.11)-(4.12) together with (B.1) we have
\begin{equation}
u_{a;b}K^{a}K^{b} = - \frac{1}{F} \pm \frac{D_{,r}}{D} \frac{V}{R}  ,
\end{equation}
and
\begin{eqnarray}
u_{a;bc}K^{a}K^{b}K^{c} = \frac{1}{D} \left( \frac{1}{F} \right)^{\cdot}
+ \frac{2}{F^2} \mp \frac{4}{F} \left( \frac{D_{,r}}{D} \right)
\frac{V}{R} + \left( \frac{D_{,r}}{D} \right)^2 \frac{V^2}{R^2} \nonumber \\
\mp \left( \frac{D_{,r}}{D} \right)_{,t} \frac{V}{DR} - \left( \frac{D_{,r}}{D}
\right)_{,r} \frac{V^2}{R^2} - \frac{D_{,r}}{D} \frac{V_{,r}V}{R^2}  ,
\end{eqnarray}
where
\begin{equation}
\frac{D_{,r}}{D} = \frac{ \left( \frac{V}{R} \right)^{\cdot}_{,r}}{ \left(
\frac{V}{R} \right)^{\cdot}} - \frac{ \left( \frac{V}{R} \right)^{\cdot}}
{ \left( \frac{V}{R} \right)}  ,
\end{equation}
and the first signs refer to outgoing rays and the second to incoming rays
( $\frac{k^{r}}{k^{t}} = \frac{\frac{dr}{ds}}{\frac{dt}{ds}} = \frac{dr}{dt}$
so "+" for outgoing and "-" for incoming rays respectively).

\pagebreak

\end{document}